\documentclass[11pt,twoside]{article}
\usepackage[dvips]{graphics}
\usepackage{star2000}
\newcommand{\oversim}[2]{\protect{\mbox{\lower0.5ex\vbox{%
   \baselineskip=0pt\lineskip=0.2ex
   \ialign{$\mathsurround=0pt #1\hfil##\hfil$\crcr#2\crcr\sim\crcr}}}}} 
\newcommand{\simgreat}{\mbox{$\,\mathrel{\mathpalette\oversim>}\,$}} % >~ sign
\newcommand{\simless} {\mbox{$\,\mathrel{\mathpalette\oversim<}\,$}} % <~ sign
\def\edcomment#1{\iffalse\marginpar{\raggedright\sl#1\/}\else\relax\fi}
\reversemarginpar

\begin{document}
\title{The Local Stellar Initial Mass Function}
 \author{Pavel Kroupa} 

\affil{Institut f\"ur Theoretische Physik und Astrophysik\\
Universit\"at Kiel, D-24098 Kiel, Germany}

%\affil{Institut f\"ur Theoretische Astrophysik\\
%Tiergartenstr.~15, D-69121 Heidelberg, Germany}

\begin{abstract}
This contribution describes the difficult task of inferring the IMF
from local star-count data, by discussing the mass--luminosity
relation, unresolved binary, triple and quadruple systems, abundance
and age spreads and Galactic structure, all of which must be accounted
for properly for the results to be meaningful. A consensus emerges
that the local IMF may be represented by a two-part power-law, with
indices $\alpha=1-1.5$ for stars with mass $m\simless 0.5\,M_\odot$,
and the Salpeter value $\alpha=2.3$ for more massive stars, but some
uncertainties remain. Notable is also that the sensitivity of the
stellar luminosity function (LF) to the derivative of the
mass--luminosity relation is very evident in the (local) Hipparcos and
HST, open-cluster and globular-cluster LFs, thus allowing tests of
stellar structure theory.  The upcoming astrometry space missions DIVA
and GAIA will undoubtedly lead to significant advances in this field.
\end{abstract}


%======================================================================
\section{Introduction}
\noindent
The distribution of masses of stars born together, the {\it initial
mass function} (IMF), determines the appearance and evolution of
galaxies and star clusters, and composes the boundary condition for
star-formation when it ends.  Any affirmed variation of the IMF with
local conditions would pose tremendously important constraints on our
understanding of how stars form.  The detailed shape of the IMF, let
alone any variation of it between populations, has been arduous to
distill from the observational data, with a large amount of serious
work remaining to be done in this fundamentally important field.

This text focuses on the derivation of the IMF from
solar-neighbourhood (sn) star-count data.  The sn constitutes a
particularly important stellar sample because of it's proximity, and
because it represents a mixture of many star-formation (sf)
events. The {\it solar-neighbourhood IMF} is therefore an {\it average
IMF}, being valid for the on average about~5~Gyr old Milky-Way (MW)
disk stars.  Furthermore, by studying the sn, methods can be developed
that are of general use for the interpretation of more distant stellar
populations.

Because of the restriction to the sn, the mass range covered by this
contribution spans $0.1-3\,M_\odot$, more massive stars being too
rare. Very-low-mass stars and brown dwarfs are dealt with by Chabrier
\& Baraffe (2000), while Massey (1998) addresses the IMF for massive
stars. Empirical evidence for systematic variations over all stellar
masses ($10^{-2}-10^2\,M_\odot$) is studied by Kroupa (2000a), and
Meyer et al. (1999) review the IMF in star-forming regions.

%======================================================================
\section{From the LF to the IMF}
\noindent
The mass, $m$, of an isolated main-sequence star can be determined
from its absolute luminosity, $l$, and the mass--luminosity relation,
$m = m(l,\tau,{\rm [M/H]},\mbox{\boldmath$s$})$, where $\tau$, [M/H]
and {\boldmath$s$} are, respectively, the age, metal abundance and
rotational angular momentum vector (spin).  The corresponding
mass--(absolute-magnitude) relation is $m(M_P)$, where $P$ represents
some photometric pass band, and the other parameters have been dropped
for conciseness.

The number of stars in the absolute-magnitude interval $M_P+dM_P$ to
$M_P$ {\it and} in the corresponding mass interval $m$ to $m+dm$ is
$dN=-\Psi\,dM_P=\Xi\,dm$, where $\Xi(m)$ is the present-day mass
function (PDMF).  The distribution of main sequence stars by
luminosity, $\Psi(M_P)$ (the present-day stellar luminosity function,
PDLF), is therefore related to the PDMF through
\begin{equation}
\Psi(M_P) = - \Xi(m)\,{dm\over dM_P}.
\label{eqn:psi}
\end{equation}
Star-counts allow an estimate of $\Psi$ from which $\Xi$ can be
derived, assuming $m(M_P,\tau,\left[{\rm
M/H}\right],\mbox{\boldmath$s$})$ is known.

To infer the IMF, $\xi(m)$, corrections for stellar evolution on and
off the main sequence are necessary.  As a star ages while burning
hydrogen in the core, its effective temperature increases slightly
owing to core contraction and its luminosity increases.  A rotating
star has a smaller internal pressure than a non-rotating star owing to
the centrifugal force, leading to reduced internal temperatures, lower
thermonuclear power, and subluminous stars. Models indicate that the
luminosity may be reduced by about 10~per cent for rigidly rotating
stars (e.g. Mendes, D'Antona \& Mazzitelli 1999; Sills, Pinsonneault
\& Terndrup 2000).  Main-sequence stars with a spectral type later
than approximately F4 are slow rotators, whereas more massive stars
are fast rotators. This is a result from mass loss and magnetic
activity, causing the stars with convective envelopes (spectral type
F4 and later) to loose spin angular momentum.  Corrections for the
time-evolution of {\boldmath$s$} should therefore also be applied.
Mass-loss during the main-sequence phase of stars adds another
complication, as it is mass and time dependent.  Finally, stars of
different chemical composition have different effective temperatures
and luminosities.

Such effects lead to a widening of the main sequence.  Explicit
corrections for each of these effects are difficult, and in practice
an average, or empirical $m(M_P)$ relation is often adopted. But in
doing this, subtle but important features in the $m(M_P)$ relation may
be lost (e.g. Belikov et al. 1998 for fine structure in the Pleiades
LF).  Additionally, $\Psi$ must be corrected for hidden companion
stars, because there is not a one-to-one mapping between a system's
luminosity and it's mass.

%------------------------------------------------------------------
\section{The LF}
\label{sec:lf}
From the many attempts of constructing $\Psi$ (see Scalo 1986 for a
comprehensive review), only two withstood the test of time in the
sense that their respective systematic biases can be handled
readily. These two fundamentally different but complementary
approaches rely on constructing complete stellar samples from
trigonometric and photometric parallax surveys.  The former requires
all sample-stars to have high-quality trigonometric parallax
measurements, and is thus confined to a very small volume around the
Sun and mostly the northern hemisphere (for historical reasons), whereas
the latter extends to much larger distances through sensitive,
pencil-beam surveys.

\vskip 1.5mm
\noindent{\bf \underline{The nearby or single star LF:}} ($\Psi_{\rm
near}$) Ground-based measurements can only measure {\it relative
trigonometric parallaxes}, and suffer from systematic errors, because
they rely on measuring the parallactic shift relative to many
background stars that must lie typically less than 1~degree from the
target star to avoid refractive atmospheric effects.  An astrometry
satellite, on the other hand, allows a reference star to be chosen for
each target star, such that the reference star is aligned along the
Earth-Sun axis at the times of measurement. The reference star thus
remains fixed on the celestial sphere while the target star shows the
maximum parallactic shift, if the angle between the target and
reference star is~90 degrees. Space astrometry thus allows the
measurement of {\it absolute trigonometric parallaxes} (e.g. Perryman
et al. 1995).

Distance measurements from space thus lead to a significantly improved
estimate of the LF, because trigonometric-parallax-limited surveys are
biased. The bias is larger for a larger uncertainty in distance
measurements: The number of stars per radial shell increases as
$\propto r^2$. Thus, there are more stars lying just outside the
survey distance limit but that have a distance error that places them
(in the measurement) within the distance limit, than the number of
stars that are within the distance limit but are measured to be
outside. The result is that less-accurate measurements
(i.e. ground-based parallaxes) overestimate $\Psi_{\rm near}$, and
that the average luminosities are underestimated (Lutz \& Kelker 1973;
Smith \& Eichhorn 1996; Oudmaijer, Groenewegen \& Schrijver 1998).

\begin{table}
{\tiny
\begin{minipage}[t]{5cm}
\hspace{1cm}
  \begin{tabular}{*{6}{c}}
%   \tableline\tableline
    $M_V$ & $\Psi_{\rm near}\times10^3$ & $\delta \Psi_{\rm
    near}\times10^3$ & $N$ &$V$ &$r_{\rm compl}$\\
%    \tableline
      &{\scriptsize (stars/pc$^3$/mag)}
        &{\scriptsize (stars/pc$^3$/mag)}
        &    &{\scriptsize (pc$^3$)} &{\scriptsize (pc)}\\
%    \tableline\tableline
$-1$     &0.015    &0.015     &1     &  65450        & 25  \\
$+0$     &0.092    &0.038     &6     &  65450        & 25  \\    
$+1$      &0.24     &0.060     &16    &  65450        & 25  \\
$+2$      &0.41     &0.079     &27    &  65450        & 25  \\
$+3$      &1.10     &0.13      &72    &  65450        & 25  \\
$+4$      &1.59     &0.16      &104   &  65450        & 25  \\
$+5$      &2.92     &0.21      &191   &  65450        & 25  \\
$+6$      &2.98     &0.21      &195   &  65450        & 25  \\
$+7$      &2.92     &0.21      &191   &  65450        & 25  \\
$+8$      &3.34     &0.26      &164   &  49115        & 25  \\
$+9$      &4.18     &0.41      &105   &  25147        & 20  \\    
$+10$     &7.00     &1.49      &22    &   3143        & 10  \\
% ground-based:
$+11$      &10.2    &1.8       &32    &    395        &5.2  \\
$+12$      &17.7    &6.7       &7     &    395        &5.2  \\
$+13$      &12.7    &5.7       &5     &    395        &5.2  \\
$+14.5$    &13.9    &4.2       &11    &    395        &5.2  \\
$+16.5$    &11.4    &3.8       &9     &    395        &5.2  \\
%
%    \tableline\tableline
\end{tabular}
\end{minipage}
}
\caption{ {\small The nearby, main-sequence PDLF estimated using
Hipparcos trigonometric parallax data. For $M_V\le8.5$, all
declinations are used, but for $M_V>8.5$, $\delta\ge-30$ (Jahreiss \&
Wielen 1997). For $M_V\ge10.5$, $\Psi$ relies on trigonometric
parallax determination from the ground (Kroupa 1998, and references
therein) using $r_{\rm compl}=5.20$~pc and $\delta>-20^{\rm o}$,
giving a volume of~395~pc$^3$. For each magnitude bin, the actual
number of stars in the survey volume is $N$. The last four bins have
been combined to two 2-magnitude wide bins.  }}
\label{tab:lfhip}
\end{table}

The result of the Hipparcos mission (e.g. Perryman et al. 1997) is a
re-derivation of $\Psi$ for $M_V\simless 11$, such that $\Psi_{\rm
Hip}<\Psi_{\rm old}$ (Wielen, Jahreiss \& Kr\"uger 1983) by about
15~per cent.  This is a direct consequence of the above mentioned
Lutz-Kelker bias.  Table~\ref{tab:lfhip} contains the PDLF estimated
from Hipparcos data for $M_V<10.5$, extended to fainter magnitudes
using ground-based parallax data.  Most stars have been scrutinised in
much detail (e.g. Duquennoy \& Mayor 1991; Fischer \& Marcy 1992), so
that virtually all components in multiple systems are known and
counted individually.

For fainter stars ($M_V\simgreat 11$), $\Psi_{\rm near}$ remains
defined by the sample of stars with ground-based
parallaxes. Completeness of the survey volume ends near $r_{\rm
compl}=5$~pc (Jahreiss 1994; Henry et al. 1997).  The claim that the
completeness limit can be extended to beyond 8~pc for M~dwarfs (Reid
\& Gizis 1997), based on including photometric parallax estimates that
allow systems with much larger distances to enter the sample
(Section~\ref{sec:compl}), is thus unlikely to be correct (Chabrier \&
Baraffe 2000). This is evident by virtue of long-term radial velocity
surveys uncovering previously unknown binary systems {\it to known
primaries} within $5<r<9$~pc and declinations $\delta>-16^{\rm o}$
(Delfosse et al. 1999).

\vskip 1.5mm
\noindent{\bf \underline{The photometric or system LF:}} ($\Psi_{\rm
phot}$) $\Psi_{\rm near}$ is poorly defined for $M_V>11$. Other
techniques for estimating the LF of faint main sequence stars were
consequently developed.  A major driving force in this endeavour was
the attempt to quantify how much mass is ``hidden'' in the faintest
stars, given that some investigations prior-to and during~1980 arrived
at significant amounts of dark matter apparently distributed like the
Galactic disk (e.g. Bahcall 1984), which is now definitely known not
to be the case (Cr\'ez\'e et al. 1998).

This approach, made possible by the advent of automatic plate
measuring machines and pioneered by Reid \& Gilmore (1982), involves
deep photographic or CCD imaging in two or three photometric pass
bands. The solid angle of such a survey is small, but the volume
surveyed is large if the survey extends to $r_{\rm compl}\simgreat
100$~pc.  Stellar distances, and thus volume number densities, are
derived by using photometric parallax. Interstellar absorption is
negligible within a few hundred~pc if the field of view is directed
out of the Galactic disk. The distance limit, $r_{\rm compl,ph}$, to
which the survey is complete, decreases with increasing $M_P$, and can
be calculated given that the flux limit below which the survey becomes
incomplete is known. Similar to the Lutz-Kelker bias, the {\it
Malmquist bias} distorts the shape of the flux-limited $\Psi_{\rm
phot}$: an observer overestimates the number of stars, which are
intrinsically brighter than the average star of a given colour. This
bias arises because a colour does not uniquely specify the absolute
magnitude of a star, which also depends on age, metallicity, spin and
multiplicity. This bias can be corrected for (Stobie, Ishida \&
Peacock 1989), and only the Malmquist-corrected photometric LF,
$\Psi_{\rm phot}$, is considered from here on.

While $\Psi_{\rm near}$ can only be defined with one sample, the many
possible line-of-sights out of the Galactic disk allow many
independent estimates of $\Psi_{\rm phot}$.  A consistent finding
among these surveys, each yielding about 30--60~stars with
$M_V\simgreat11$, is that $\Psi_{\rm phot}$ has a maximum at
$M_V\approx12$ with a decline at fainter magnitudes.  A weighted
average, corrected to the Galactic-disk midplane density,
$\overline{\Psi}_{\rm phot}$, was calculated from the individual
surveys (Kroupa 1995a).

The Hubble Space Telescope (HST) delivers diffraction-limited images
down to the flux limit ($I\approx24$).  Contamination through galaxies
is thus essentially removed, and the stellar distribution is probed to
distances of a few kpc. For example, an M~dwarf with $M_V=16$ has
$M_I=12.1$, approximately, and is thus detectable to a distance of
about 2.4~kpc.  Malmquist bias is negligible, because, in a
field-of-view that is directed at a Galactic latitude northward of
about 45~deg, the flux limit essentially corresponds to a true volume
limit because the photometric distance limit lies well outside the
stellar distribution.  Many fields are combined to yield reasonable
statistics. The Groth-strip survey (Gould, Bahcall \& Flynn 1997) adds
about 45~stars with $12.5<M_V<16.25$.  The HST results confirm the
finding arrived at from the ground.

\begin{figure}
\plotfiddle{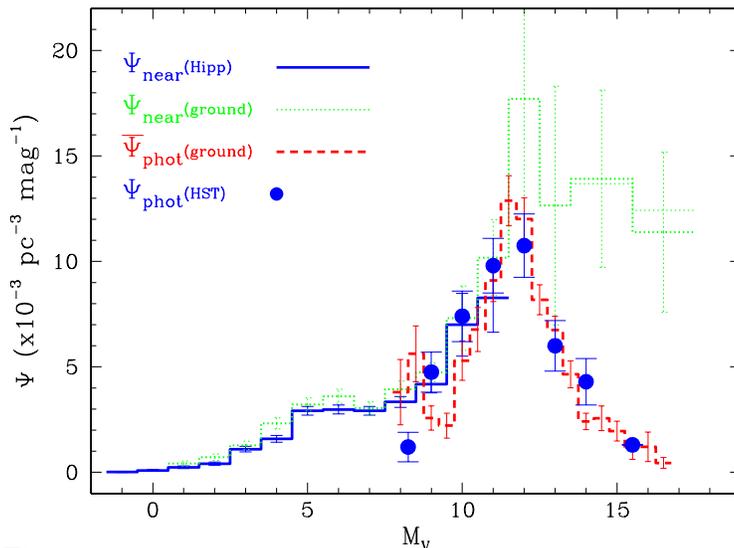}{6.1cm}{-90}{38}{38}{-170}{210}
\caption{\small The photometric LF corrected for Malmquist bias and at
the midplane of the Galactic disk ($\Psi_{\rm phot}$) is compared with
the nearby LF ($\Psi_{\rm near}$). The average, ground-based
$\overline{\Psi}_{\rm phot}$ (dashed histogram, Kroupa 1995a) is
confirmed by HST data (solid dots, Gould et al. 1997). The
ground-based trigonometric-parallax sample (dotted histogram)
systematically overestimates $\Psi_{\rm near}$ due to the Lutz-Kelker
bias, thus lying above the improved estimate provided by Hipparcos
data (solid histogram, Table~1). The thin dotted histogram at the
faint end indicates the level of refinement provided by recent stellar
additions (see Kroupa 1998 and references therein).  }
\label{fig:lfn_nearby}
\end{figure}

\vskip 1.5mm
\noindent{\bf \underline{Nearby versus photometric LF:}}
The resulting LFs are plotted in Fig.~\ref{fig:lfn_nearby}.
%-------------------------------------
% Figs: /PAPERS/Habil/Vorl/ProgsaFigs/lfn_cf_star2000.sm
%-------------------------------------
$\Psi_{\rm near}$ increases by an order of magnitude from
$M_V\approx0$ to $M_V\approx17$. Approximately 70~per cent of all
stars have $M_V>10.5$, demonstrating the possible importance of faint
main sequence stars for the mass-budget of a galaxy and star cluster.\
$\Psi_{\rm near}$ shows interesting structure. It increases
monotonically with increasing $M_V$ until $M_V\approx5$. It is flat in
the interval $5.5<M_V<8.5$, the {\it Wielen dip}, but continues to
rise again until $M_V\approx12$ (Section~\ref{sec:mlr}). Poisson
uncertainties are too large at fainter $M_V$ to allow firm conclusions
about the shape of the LF, but it is clear that $\Psi_{\rm near}$
flattens for $M_V>12.5$. An important point to remember is that any
structure in $\Psi_{\rm near}$ and $\Psi_{\rm phot}$ is smeared out
because of the spread in metallicities, ages, spins and distance
errors (correction for Malmquist bias removes this partially in
$\Psi_{\rm phot}$).

The comparison of $\Psi_{\rm near}$ and $\Psi_{\rm phot}$ shows that
the two are very different and appear to measure different stellar
populations for $M_V>13$.  Thus, while 20~stars are counted with
$13.5<M_V<16.5$ in $\Psi_{\rm near}$, only 2~stars are seen in
$\overline{\Psi}_{\rm phot}$ within a volume of~395~pc. The majority
of faint stars in the solar neighbourhood have ages $\tau>1$~Gyr, and
the one-dimensional velocity dispersion is about~30~km/s (Meusinger,
Reimann \& Stecklum 1991; KTG93; Reid, Hawley \& Gizis 1995).  This
means that essentially all stars located within a spherical volume
with a diameter of~600~pc will have been replaced within~20~Myr. The
observed difference is not a local over density.

Reid \& Gizis (1997) suggest that the $M_V(V-I)$ relation used for
photometric parallax estimation is non-linear near $M_V=12$, which has
not been taken into account by previous work (e.g. Stobie et al. 1989;
KTG93; Gould et al. 1997). The effect is such that previous work
underestimates $M_V$, which leads to underestimates in the density
near $M_V=12$, since distance estimates are too large. Reid \& Gizis
attribute the apparent difference between $\Psi_{\rm near}$ and
$\Psi_{\rm phot}$ to this error.

\begin{figure}
\plotfiddle{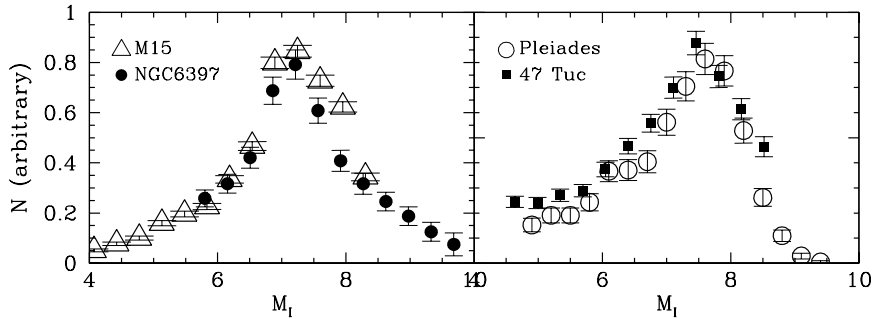}{4cm}{-90}{45}{45}{-170}{250}
\caption{\small $I$-band LFs of stellar {\it systems} in four star
clusters (M15: de Marchi \& Paresce 1995a, distance modulus $\Delta
m=m-M=15.25$~mag; NGC~6397: Paresce, de Marchi \& Romaniello 1995,
$\Delta m=12.2$; Pleiades: Hambly, Jameson \& Hawkins 1991, $\Delta
m=5.48$; 47~Tuc: de Marchi \& Paresce 1995b, $\Delta m=13.35$).  }
\label{fig:lfn_cf}
\end{figure}

The shape and amplitude of the maximum in $\Psi_{\rm phot}$ may thus
require some revision. However, notable is that system LFs for a wide
variety of star clusters also show a {\it very pronounced maximum}
near $M_V=12$ (Fig.~\ref{fig:lfn_cf}) with very similar shape, which
is not surprising if this structure is due to the derivative of the
$m(M_P)$ relation in a 'pure' population, these LFs not being mired by
age, metallicity and distance spreads (different spins may have an
influence in young clusters).  Furthermore, consulting the $M_V(V-I)$
data plotted by Baraffe et al. (1998, fig.~5), the degree of
non-linearity in this relation promoted by Reid \& Gizis is not
evident.
%-------------------------------------
% Figs: /PAPERS/Habil/Vorl/ProgsaFigs/cf_LFs_star2000.sm
%-------------------------------------

%-------------------------------------------------------------
\section{The mass--luminosity relation}
\label{sec:mlr}
Eqn.~\ref{eqn:psi} shows that any non-linear structure in this relation
is mapped into observable structure in the LF, provided the MF does
not have compensating structure. Such a conspiracy is very unlikely
because the MF is defined through the star-formation process, but the
$m(M_P)$ relation is a result of the internal constitution of
stars. 

A thorough understanding of $dm/dM_P\,(M_P)$ is thus necessary to
avoid unphysical features entering the IMF.  Much effort has gone into
establishing high-quality observational constraints from binary stars
(Popper 1980; Andersen 1991; Henry \& McCarthy 1993; Malkov, Piskunov
\& Shpil'kina 1997; Henry et al. 1999).  The $m(M_V)$ relation for
main-sequence stars is shown in fig.~1 in KTG93.  It is immediately
apparent that the slope is very small at faint luminosities.  This
holds true in other photometric passbands as well (Henry \& McCarthy
1993) and leads to large uncertainties in the MF near the hydrogen
burning mass limit.

The observational data (Andersen 1991) show that the
log$_{10}[m(M_V)]$ relation is essentially linear for
$m>2\,M_\odot$. However, as the mass of a star is reduced, and for
effective temperatures between~6000 and~7000~K, H$^-$ opacity becomes
increasingly important through the short-lived capture of electrons by
H-atoms. For stars of spectral type~F5 to~G, H$^-$ provides more
than~60~per cent of the continuous opacity. This results in reduced
stellar luminosities for intermediate and low-mass stars. The $m(M_V)$
relation becomes less steep in the broad interval $3<M_V<8$ (fig.~1 in
KTG93), leading to the Wielen dip evident in $\Psi_{\rm near}$
(Mazzitelli 1972; Meusinger 1983; D'Antona \& Mazzitelli 1986; Kroupa,
Tout \& Gilmore 1990, KTG90; Haywood 1994).

The modern data (Henry et al. 1993; fig.~2 in Kroupa 1998) confirm the
steepening in the interval $10<M_V<13$ postulated by KTG90 to be the
origin of the maximum in $\Psi_{\rm phot}$ near $M_V=12$.  The
$m(M_V)$ relation steepens near $M_V=10$ because the formation of
H$_2$ in the outer shells of main-sequence stars causes the mean
molecular weight to be larger for less massive stars, invoking core
contraction. This leads to brighter luminosities and full convection
for $m\le0.35\,M_\odot$.  The $m(M_V)$ relation flattens again for
$M_V>14$, $m<0.2\,M_\odot$, as degeneracy in the stellar core becomes
increasingly important for smaller masses, thus supporting the core
against further contraction (Chabrier \& Baraffe 1997).

A pronounced local maximum in $-dm/dM_V(M_V)$ results at
$M_V\approx11.5$. Artificial suppression of H$_2$ formation eliminates
this maximum (fig.~3 in KTG90).  Different theoretical $m(M_P)$
relations have extrema in $dm/dM_P(M_P)$ at different $M_P$,
suggesting the possibility of testing stellar structure theory near
the critical $m\approx0.35\,M_\odot$, where stars become fully
convective (Kroupa \& Tout 1997; Brocato, Cassisi \& Castellani 1998).

%--------------------------------------------------------------
\section{Additional complications}
\label{sec:compl}
\vskip 1.5mm
\noindent{\bf \underline{Multiple stellar systems:}} In addition to
the non-linearities in the $m(M_P)$ relation, unresolved multiple
systems affect the MF derived from an observed LF, in particular since
{\it no stellar population is known to exist that has a binary
proportion smaller than 50~per cent}, apart from the dynamically
highly evolved globular clusters (e.g. Kroupa 2000b and references
therein).

Suppose an observes sees 100~systems. Of these~40, 15~and 5~are
binary, triple and quadruple, respectively. There are thus
85~companion stars which the observer is not aware of if none of the
multiple systems are resolved. Since the distribution of secondary
masses for a given primary mass is not uniform, but typically
increases with decreasing mass (e.g. Malkov \& Zinnecker 2000), the
bias is such that low-mass stars are significantly underrepresented in
any survey that does not detect companions (Fig.~\ref{fig:mods} below;
Piskunov \& Malkov 1991; Kroupa, Tout \& Gilmore 1991, KTG91; KTG93;
Holtzman et al. 1997; 1998).

Counting a multiple system as a single star has the following effects
in a photometric survey: (i) if the companion(s) are bright enough to
affect the system luminosity noticeably, then the estimated
photometric distance will be too small, and (ii) the companions are
lost from the star-count analysis. The former effect enhances the
apparent stellar number density at brighter magnitudes.  This is
evident in fig.~7 in KTG91, where the system LF lies above the
single-star LF for $M_V<10$. However, in reality this is countered by
the larger effective photometric distance limit together with the
approximately exponential stellar density fall-off perpendicular to
the Galactic plane, implying that the photometric LF {\it not}
corrected for Malmquist bias is about equal to or smaller than
$\Psi_{\rm near}$ for most $M_V$ (KTG93).  The latter effect~(ii)
reduces the star-counts at faint magnitudes leading to a significant
bias, because a G-, K- and bright M-dwarf has, on average, one or more
faint M-dwarf companions.  Note that a faint companion will also be
missed if the system is formally resolved but the companion lies below
the flux limit of the survey.

The true distance limit can be significantly larger than the nominal
value when estimating a photometric parallax. If a stellar system in
the star-count survey is composed of two equal-mass stars, it has an
$M_V$ brighter by 0.75~mag than a single star of the same colour. If,
in addition, the system has a metallicity such that the combined
absolute magnitude is $3\sigma_{M_V}=1.5$ brighter than a star of the
same colour (where $\sigma_{M_V}$ is the spread in $M_V$ for a given
colour, the 'cosmic scatter'), then the resulting absolute magnitude
of the unresolved system can be brighter by as much as $\delta
M_V=2.25$, which implies that it can be seen $10^{\delta M_V/5}=2.82$
times as far as the nominal distance limit. For example, if the
nominal distance limit is 130~pc, then systems as far away as 366~pc
can in principle enter the sample.  While this is an extreme case, it
does demonstrate that photometric star-count surveys are contaminated
by systems that are beyond the nominal distance limit, which, of
course, needs to be part of any model (KTG93).

\vskip 1.5mm
\noindent{\bf \underline{Metallicity and Ages spreads:}} The
metallicity distribution, which results from the chemical evolution of
the Galactic disc (see e.g. Gilmore \& Wyse 1991; Samland, Hensler \&
Theis 1997; Tsujimoto et al. 1997; Rocha-Pinto et al. 2000), is
non-Gaussian with a mean near [Fe/H]$\approx -0.2$~dex, a width of
roughly~0.3~dex, and a tail towards low iron abundance. K~dwarfs have
a similar distribution as G~dwarfs (Flynn \& Morell 1997; Rocha-Pinto
\& Maciel 1998). A somewhat different abundance distribution for stars
of different mass is to be expected though because of the
age--metallicity relation (e.g. Meusinger et al. 1991; Ng \& Bertellie
1998; Carraro, Ng \& Portinari 1998). Stars with short life-times
usually only sample the high-metallicity range.

The distribution of stellar ages is a major source of uncertainty in
the analysis of counts of stars that have a mass in the range
$0.8\,M_\odot-3\,M_\odot$ (Haywood, Robin \& Cr\'ez\'e 1997a; Maciel
\& Rocha-Pinto 1998). The {\it SFR--$\alpha$ degeneracy} is emphasised
by Binney, Dehnen \& Bertelli (2000).  Such stars evolve along and off
the main sequence within the age of the Galactic disc ($9-12$~Gyr), so
that the distribution of luminosities and colours for main-sequence
stars with the same mass depend critically on the sf rate (SFR). Stars
with $m\simgreat3\,M_\odot$ have life-times much shorter than the age
of the Galactic disc, and consequently map only the most recent sf
history.  Only the ratio of the average SFR to the present SFR is
important in adjusting star counts of these stars to the number of
stars with $m\simless0.8\,M_\odot$, which amount to {\it all} stars
ever formed in the Galactic disc, assuming $\xi(m)$ is continuous
across $m\approx 3\,M_\odot$ and unchanging.

There are tentative hints at non-uniform star-formation histories.
For example, Noh \& Scalo (1990) discuss a marginal feature in the
white-dwarf-luminosity function which may suggest a burst of star
formation about $3\times10^8$~yr ago. This burst may have had a
duration of $\simless10^8$~yr, and may have contributed about 10~per
cent of the stars in the solar neighbourhood.  However, only stars
with $m\simless0.2\,M_\odot$, that have long pre-main sequence
contraction times, appear brighter by $\delta M_V\simgreat0.2$~mag if
they are younger than about $3\times10^8$~yr old (Baraffe et
al. 1998). Possible bursts of the SFR about 8~and 3~Gyr ago are
suggested by Rocha-Pinto \& Maciel (1997), Rocha-Pinto et al. (2000),
and Hernandez, Valls-Gabaud \& Gilmore (2000) discuss the local sf
history during the past 3~Gyr evident from Hipparcos data. 

In summary, fully consistent modelling of star-count surveys must
include models of the age distribution of Galactic field stars, unless
only the restricted mass range $0.2\simless m/M_\odot \simless 0.8$ is
studied.

\vskip 1.5mm
\noindent{\bf \underline{Galactic structure:}} Star count surveys used
to estimate the local LF of late-type main-sequence stars are
restricted to distances of less than about one~kpc. Galactic
components such as the bulge or the stellar halo are thus not very
important. A summary of Galactic structure and related topics may be
found in Gilmore, Wyse \& Kuijken (1989).

The structure of the Galactic disc can be described reasonably well by
exponential density distributions in radial and vertical
directions. The radial scale length is about 2.5~kpc (Bienaym\'e \&
S\'echaud 1997; Porcel et al. 1998).  The vertical structure can be
fitted by two exponential distributions: the normal disc with a scale
height of $h\approx250$~pc (KTG93; Haywood, Robin \& Cr\'ez\'e 1997b;
M\'endez \& Guzm\'an 1998), i.e. {\it not} 300--350~pc, and the thick
disc with a scale height of about 1~kpc.  At the Galactic midplane,
the thick disc contributes only a few per cent to the number density
of disc stars (Vallenari, Bertelli \& Schmidtobreick 2000), and the
stellar halo contributes about 0.1~per cent (Robin \& Cr\'ez\'e 1986).
Finally, the Sun appears to be located about 10--20~pc ``above'' the
Galactic plane (Marsakov \& Shevelev 1995; Reed 1997; Minezaki et
al. 1998).  This off-set can usually be neglected in the analysis of
star-counts, but may induce some anisotropy in $\Psi_{\rm near}$. Deep
photometric surveys, however, have to be corrected for Galactic
structure.

The contamination of $\Psi_{\rm phot}$ with thick-disk stars can be
significant. These are, on average, metal-poorer than 'normal' disk
stars that make-up $\Psi_{\rm near}$, so that photometric parallax
would lead to systematically wrong distance estimates and thus space
densities, since an inappropriate colour--magnitude relation is used.
In a survey with distance limit $z_o$ covering a solid angle $\Omega$,
the number of stars detected is
$N=\Omega\int_0^{z_o}\rho(z)z^2dz$. The density fall-off perpendicular
to the Galactic disk can be approximated by $\rho(z)=\rho_o\,{\rm
exp}(-z/h)$, $h$ being the scale height. Thus
$N=\Omega\rho_o\,h^3\,[2-(2+2y+y^2)/{\rm exp}(y)]$, $y\equiv z_o/h$,
and about $1/3$ of all stars in an HST survey with $z_o=1$~kpc are
thick-disk stars.  The excellent agreement between the ground-based
and HST $\Psi_{\rm phot}$, evident in Fig.~\ref{fig:lfn_nearby}, is,
however, in-line with the general finding that the shape of the system
LF, i.e. of $dm/dM_P$, is not very sensitive of the metallicity of the
underlying population (Fig.~\ref{fig:lfn_cf} and Kroupa \& Tout 1997).

%-----------------------------------------------------
\section{The IMF}
The above sections give an impression of the complexity required to
analyse local star-counts in order to infer the underlying IMF. Such
complex models have been constructed by KTG93, with the finding that
$\alpha_1=1.3\pm0.5$ for $m<0.5\,M_\odot$ and $\alpha_2\approx2.2$ for
$m>0.5\,M_\odot$ ($\xi(m)\propto m^{-\alpha_i}$).  This is supported
by the dynamical population synthesis model of the sn presented by
Kroupa (1995b). This model does not rely on random association of
stars into binaries, and nicely reproduces the empirical mass-ratio
distribution of late-type binaries of Reid \& Gizis (1997) (fig.~1 in
Kroupa 2000b). Using an entirely different empirical $m(M_V)$ relation
but a simpler star-count model, leads to the same result
(KTG91). These models always aim at reproducing $\Psi_{\rm near}$ and
$\Psi_{\rm phot}$ simultaneously, thereby improving the constraints.
A slightly shallower IMF ($\alpha=1.05$, $m\simless1\,M_\odot$) is
arrived at by Reid \& Gizis (1997), although their nearby star-count
sample is incomplete (Section~\ref{sec:lf}), and stellar evolution is
not modelled.  Gould et al. (1997) find $\alpha_1\approx0.9$
($m<0.6\,M_\odot$) and $\alpha_2\approx2.2$ ($m>0.6\,M_\odot$) by
analysing their $\Psi_{\rm phot}$ constructed from HST data, but they
apply only very crude corrections for unresolved binaries and also do
not include stellar evolution.  A somewhat steeper local IMF
($\alpha_1=2\pm0.5$) is arrived at using a theoretical $m(M_V)$
relation and $\Psi_{\rm near}$ (M\'era, Chabrier \& Baraffe
1996). This $m(M_V)$ relation is expected to be further revised as
additional opacity sources in the optical are introduced (Baraffe et
al. 1998).  Finally, the very extensive modelling of the sn by Haywood
et al. (1994; 1997a; 1997b) arrives at $\alpha\approx1.7$ for
$m<1\,M_\odot$ and $\alpha\approx2$ for $1-3\,M_\odot$.  These models
incorporate stellar evolution, unresolved binaries and Galactic-disk
structure, but again only rely on $\Psi_{\rm near}$ to constrain
$\alpha$ for $m\simless0.5\,M_\odot$.

A consensus thus appears to be emerging that the MF has
$\alpha_2\approx2.3$ for $m\simgreat0.5 \,M_\odot$ and
$\alpha_1\approx1-1.5$ for $0.1<m<0.5\,M_\odot$, the flattening of
the IMF near $0.5\,M_\odot$ becoming evident when the statistically
better-defined $\Psi_{\rm phot}$ is used in addition to $\Psi_{\rm
near}$.

%-----------------------------------------------------
\section{The Future}
The remaining discrepancies for $m\simless0.5\,M_\odot$ can {\it
partially} be aleviated with significantly improved
trigonometric-parallax limited star-counts, as will become available
with the upcoming astrometry satellites DIVA (R\"oser 1999) and GAIA
(Lindegren \& Perryman 1996; Gilmore et al. 1998b). DIVA will fly for
two years in 2004, and will find {\it all systems} with
$M_V\simless16$ and with $r_{\rm compl}\simless15$~pc, measuring their
distances and luminosities .  GAIA has also been approved by the
European Space Agency.

\begin{figure}
\plotfiddle{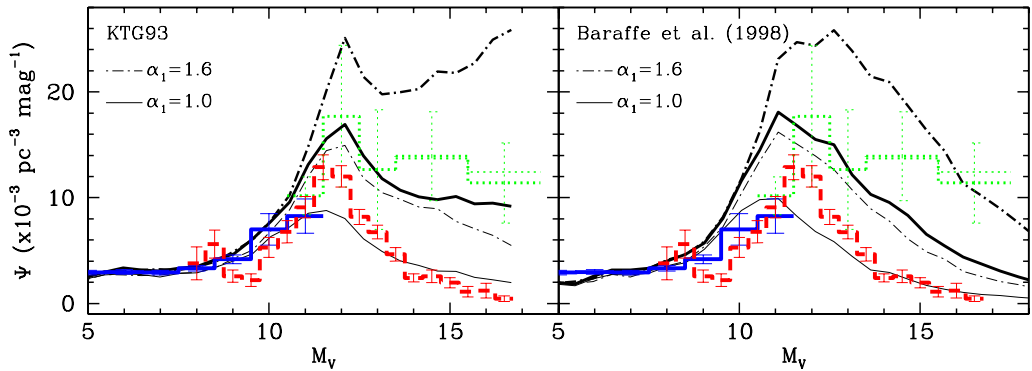}{4cm}{-90}{55}{55}{-230}{307}
\caption{\small Model LFs assuming $\alpha_1=1.6$ (dash-dotted lines)
or $\alpha_1=1.0$ (solid lines) for $0.08-0.5\,M_\odot$ and the
Salpeter value $\alpha_2=2.3$ for $0.5-1.0\,M_\odot$ ($\xi(m)\propto
m^{-\alpha_i}$).  Thick lines are the single-star LF, whereas thin
lines show the system LF (all companions merged photometrically) for a
population consisting of 8000 single stars, 8000 binaries, 3000
triples and 1000 quadruples (40:40:15:5~per cent, respectively).
Companions are combined randomly from the IMF.  The models assume
perfect photometry, no distance errors and no metallicity or age
spread. The left panel is for the semi-empirical $m(M_V)$ relation
from KTG93, and the right panel uses the theoretical relation from
Baraffe et al. (1998, 5~Gyr isochrone, [M/H]$=0$). The histograms are
as in Fig.~\ref{fig:lfn_nearby}, and the models are scaled to fit the
data near $M_V=7$ (equal scaling for both the single-star and system
LFs).  }
\label{fig:mods}
\end{figure}

Examples of the type of {\it single-star} and {\it system} LFs
expected using different $m(M_V)$ relations are shown in
Fig.~\ref{fig:mods}. 
%-------------------------------------
% Figs: /PAPERS/Imfpap4/Figs_progs/lfn_star2000.sm
%-------------------------------------

The following two warnings must be stressed at this point: 1) DIVA and
GAIA will provide exquisite data for the motions and positions of {\it
stellar systems} only, because a significant fraction of the
companions of multiple systems will again remain undetected until
follow-up observations scrutinise every new faint stellar system
discovered by the satellites, a rather daunting task that will
probably only be possible with automated survey telescopes.  2) Any
model of DIVA and GAIA data will have to include the age and
metallicity distributions, since these smear-out $dm/dM_P$-features in
the LF, as well as distance-dependent resolution of multiple systems
with the appropriate flux limits.

%==============================================================
\vskip 3mm
\noindent 
\small{I acknowledge support through DFG grant KR1635. }

%==============================================================

%==============================================================

\begin{references}
\small{
\reference Andersen, J. 1991, A\&AR, 3, 91

\reference Bahcall, J.N. 1984, \apj, 287, 926

\reference Baraffe, I., Chabrier, G.,
        Allard, F., \& Hauschildt, P. H., 1998, \aap, 337, 403

\reference Belikov, A. N., Hirte, S., Meusinger, H., Piskunov, A. E.,
	\& Schilbach, E. 1998, \aap, 332, 575

\reference Bienaym\'e, O., S\'echaud, N. 1997, \aap, 323, 781

\reference Binney, J., Dehnen, W., \& Bertelli, G. 2000, \mnras, in
	press (astro-ph/0003479)

\reference Brocato, E., Cassisi, S., \& Castellani, V. 1998, \mnras,
	295, 711

\reference Carraro, G., Ng, Y. K.,
	\& Portinari, L. 1998, \mnras, 296, 1045

\reference Chabrier, G., \& Baraffe, I. 1997, \aap, 327, 1039

\reference Chabrier, G., \& Baraffe, I. 2000, \araa, 38, in press
        (astro-ph/0006383)

\reference Cr\'ez\'e, M., Chereul, E., Bienayme, O., \& Pichon, C. 1998,
	\aap, 329, 920

\reference D'Antona, F., \& Mazzitelli, I. 1986, \aap, 162, 80

\reference Delfosse, X., Forveille, T.,
        Beuzit, J.-L., Udry, S., Mayor, M., \& Perrier, C. 1999, \aap, 
        344, 897

\reference de Marchi, G., \& Paresce, F. 1995a, \aap, 304, 202 

\reference de Marchi, G., \& Paresce, F. 1995b, \aap, 304, 211 

\reference Duquennoy, A., \& Mayor, M. 1991, A\&A, 248, 485

\reference Fischer, D. A., \& Marcy, G. W. 1992, \apj, 396, 178

\reference Flynn, C., \& Morell, O. 1997,
	\mnras, 286, 617

\reference Gilmore, G., \& Wyse, R. F. G. 1991,
	ApJ, 367, L55

\reference Gilmore, G., Wyse,
	R. F. G., \& Kuijken, K. 1989, \araa, 27, 555


\reference Gilmore, G., Perryman, M., Lindegren, L., et al. 1998b, in
        Astronomical Interferometry, ed. R.D. Reasenberg, 
        Proc. SPIE Vol. 3350, p. 541 

\reference Gould, A., Bahcall, J. N.,
        \& Flynn, C. 1997, \apj, 482, 913

\reference Hambly, N. C., Jameson, R. F., \& Hawkins, M. R. S.
	1991, \mnras, 253, 1

\reference Haywood, M., 1994, \aap, 282, 444

\reference Haywood, M., Robin,
	A. C., \& Cr\'ez\'e, M. 1997a, \aap, 320, 428

\reference Haywood, M., Robin,
	A. C., Cr\'ez\'e, M. 1997b, \aap, 320, 440

\reference Henry, T. J., \& McCarthy, D. W. 1993, \aj, 106, 773

\reference Henry, T. J., Ianna, P. A.,
        Kirkpatrick, J. D., \& Jahreiss, H. 1997, \aj, 114, 388

\reference Henry, T.\ J., Franz, O.\ G., Wasserman, L.\ H., Benedict,
	G.\ F., Shelus, P.\ J., Ianna, P.\ A., Kirkpatrick, J.\ D.\ \&
	McCarthy, D.\ W.\ 1999, \apj, 512, 864

\reference Hernandez, X., Valls-Gabaud, D., \& Gilmore, G. 2000,
	\mnras, 316, 605

\reference Holtzman, J. A., Mould, J. R.,
        Gallagher III, J. S., Watson, A. M., et al. 1997, \aj, 113, 656

\reference Holtzman, J. A., Watson, A. M.,
        Baum, W. A., Grillmair, C. J., et al. 1998, \aj, 115, 1946

\reference Jahreiss, H. 1994, in Science with Astronomical 
        Near-Infrared Sky Surveys, Epchtein N., Omont A., Burton B.,
        Persi P. (eds), Kluwer: Dordrecht, p.63

\reference Jahreiss, H., \& Wielen, R.,
	1997, in Proceedings of the Hipparcos Venice'97 Symposium, eds,
	M. A. C. Perryman, \& P. L. Bernacca, ESA SP-402

\reference Kroupa, P. 1995a, \apj, 453, 350

\reference Kroupa, P. 1995b, \apj, 453, 358

\reference Kroupa, P. 1998, in ASP Conf. Ser. 134, Brown Dwarfs and
	Extrasolar Planets, eds Rebolo R., Zapatero Osorio M.R., \&
	Martin E., p.483,

\reference Kroupa, P. 2000a, \mnras, in press (astro-ph/0009005)

\reference Kroupa, P. 2000b, in ASP Conf. Ser. Vol. xxx, The Formation 
        of Binary Stars, IAU Symp. 200, ed. H. Zinnecker \& R. Mathieu, in
        press (astro-ph/0010347)

\reference Kroupa, P., Tout, C. A. 1997, \mnras, 287, 402

\reference Kroupa, P., Tout, C. A., \& Gilmore, G. 1990, \mnras, 244,
	76 (KTG90)

\reference Kroupa, P., Tout, C. A., \& Gilmore, G. 1991, \mnras, 251,
	293 (KTG91)

\reference Kroupa, P., Tout, C. A., \& Gilmore G. 1993, \mnras, 262,
	545 (KTG93)

\reference Lindegren, L., \& Perryman, M. A. C. 1996, A\&AS, 116, 579

\reference Lutz, T. E., \& Kelker, D. H. 1973, \pasp,
        85, 573

\reference Maciel, W. J., \& Rocha-Pinto, H. J. 1998, \mnras, 299, 889

\reference Malkov, O., \& Zinnecker, H. 2000, \mnras, in press

\reference Malkov, O. Yu., Piskunov, A. E., \& Shpil'kina, D. A. 1997, 
        \aap, 320, 79

\reference Marsakov, V. A., \& Shevelev
	Yu. G. 1995, Astronomy Reports, 39, 559

\reference Massey, P.\ 1998, in ASP Conf.\ 
	Ser.\ 142, The Stellar Initial Mass Function, 
	eds G. Gilmore, \& D. Howell, p.17 

\reference Mazzitelli, I. 1972, Astrophys. \& Space Science, 17, 141

\reference Mendes, L. T. S., D'Antona, F. D., \& Mazzitelli, I. 1999,
	\aap, 341, 174

\reference M\'endez, R. A., \& Guzm\'an, R. 1998, \aap, 333, 106

\reference M\'era, D., Chabrier, G., \& Baraffe, I. 1996, \apj, 459, L87

\reference Meusinger, H. 1983, Astron. Nachr., 304, 285

\reference Meusinger, H., Reimann, H.-G., \& Stecklum, B. 1991, \aap, 245, 57

\reference Meyer, M.\ R., Adams, F.\ C., Hillenbrand, L.\ A.,
	Carpenter, J.\ M.\ \& Larson, R.\ B.\ 2000, in Protostars and
	Planets IV, Tucson: University of Arizona Press, eds
	Mannings, V., Boss, A.P., Russell, S.\ S., p.\ 121 

\reference Minezaki, T., Cohen, M.,
	Kobayashi, Y., Yoshii, Y., \& Peterson, B. A. 1998, \aj, 115, 229

\reference Ng, Y. K., Bertelli, G. 1998, \aap,
	329, 943

\reference Noh, H.-R., \& Scalo, J. 1990, \apj,
	352, 605

\reference Oudmaijer, R. D., Groenewegen,  M. A. T., \& Schrijver, H. 
	1998, \mnras, 294, L41O

\reference Paresce, F., de Marchi, G., \& Romaniello, M.\ 1995, \apj,
	440, 216

\reference Perryman, M. A. C., Lindegren,
        L., \& Kovalevsky, J., et al. 1995, \aap, 304, 69

\reference Perryman, M. A. C., Lindegren,
        L., Kovalevsky, J., et al. 1997, \aap, 323, L49

\reference Piskunov, A. E., \& Malkov, O. Yu. 1991, \aap, 247, 87

\reference Porcel, C., Garz\'on, F.,
	Jim\'enez-Vicente, J., \& Battaner, E. 1998, \aap, 330, 136

\reference Popper, D. M. 1980. \araa, 18, 115

\reference Reed, B. C. 1997, \pasp, 109, 1145

\reference Reid, N., \& Gilmore, G. 1982, \mnras, 201, 73

\reference Reid, N., Gizis, J.E. 1997, \aj,
        113, 2246

\reference Reid, N., Hawley, S. L., \& 
        Gizis, J. E. 1995, \aj, 110, 1838

\reference Robin, A., \& Cr\'ez\'e, M. 1986, \aap, 157, 71

\reference Rocha-Pinto, H. J., \& Maciel,
        W. J. 1997, \mnras, 289, 882

\reference Rocha-Pinto, H. J., \& Maciel
        W. J. 1998, \aap, 339, 791

\reference Rocha-Pinto, H.\ J., Maciel, W.\ J., Scalo, J.\ \& Flynn,
	C.\ 2000, \aap, 358, 850

\reference R\"oser, S. 1999, Reviews in Modern Astronomy, 12, 79

\reference Samland, M., Hensler, G.,
	\& Theis, Ch. 1997, \apj, 476, 544

\reference Scalo, J. M. 1986, Fundam. Cosmic Phys., 11, 1

\reference Sills, A., Pinsonneault, M. H., \& Terndrup, D. M. 2000,
	\apj, 534, 335

\reference Smith, H., \& Eichhorn, H. 1996,
        \mnras, 281, 211

\reference Stobie, R. S., Ishida, K.,
        Peacock, J. A. 1989, \mnras, 238, 709

\reference Tsujimoto, T., Yoshii, Y.,
	Nomoto, K., Matteucci, F., Thielemann, F.-K., \& Hashimoto, M.
	1997, \apj, 483, 228

\reference Vallenari, A., Bertelli, G., \& Schmidtobreick, L. 2000,
	\aap, 361, 73

%\reference Vesperini, E.\ \& Heggie, D.\ C.\ 1997, \mnras, 289, 898

\reference  Wielen, R.,
        Jahreiss, H., \& Kr{\"u}ger, R. 1983, in The Nearby Stars and
        The Stellar Luminosity Function, eds. A. G. Davis Philip,
        A. R. Upgren, IAU Colloq. No. 76, New York: Davis Press, 163

}
\end{references}
\end{document}